\documentstyle[preprint,eqsecnum,aps]{revtex}

\begin{document}
\draft
%\preprint{XXX}
\title{Mass Density Perturbations from Inflation with Thermal Dissipation}
\author{Wolung Lee and Li-Zhi Fang}

\address{Department of Physics, University of Arizona, Tucson, AZ 85721}
\date{\today}
\maketitle
\begin{abstract}

We study the power spectrum of the mass density perturbations in an inflation
scenario that includes thermal dissipation.  We show that the condition on
which the thermal fluctuations dominate the primordial density perturbations
can easily be realized even for weak dissipation, {\it i.e.}, the rate of
dissipation is less than the Hubble expansion. We find that our spectrum of
primordial density perturbations follows a power law behavior, and exhibits a
``thermodynamical'' feature -- the amplitude and power index of the spectrum
depend mainly on the thermodynamical variable $M$, the inflation energy scale.
Comparing this result with the observed temperature fluctuations of the
cosmic microwave background, we find that both amplitude and index of the
power spectrum can be fairly well fitted if $M \sim 10^{15}-10^{16}$ GeV.

\end{abstract}

\pacs{PACS number(s): 98.80.Cq, 98.80.Bp, 98.70.Vc}

\narrowtext

\section{Introduction}
\label{sec:level1}

In the past decade, there has been a number of studies on dissipative processes
associated with the inflaton decay during its evolution.  These studies have
shed light into the possible effects of the dissipative processes. For
instance, it was realized that dissipation effectively slows down the rolling
of the inflaton scalar field $\phi$ toward the true vacuum.  These processes
are capable of supporting the scenario of inflation \cite{AST,KT}. Recently,
inspired by several new developments, the problem of inflation with thermal
dissipation has attracted many re-investigations. 

The first progress is from the study of the non-equilibrium statistics of quantum
fields, which has found that, under certain conditions, it seems to be reasonable
to introduce a dissipative term (such as a friction-like term) into the equation
of motion of the scalar field $\phi$ to describe the effect of heat contact
between the $\phi$ field and a thermal bath.  These studies 
shown that the thermal dissipation and fluctuation will most likely appear 
{\it during} the inflation if the inflaton is coupled to light fields\cite{GM}.
However, to realize sufficient e-folds of inflation with thermal dissipation, 
this theory needs to introduce tens of thousands of scalar and fermion 
fields interacting with the inflaton in an {\it ad hoc} manner\cite{BGR}. 
Namely, it is still far from a realistic model. Nevertheless, this study
indicates that the condition necessary for the ``standard" reheating evolution
 -- a coupling of inflaton with light fields -- is actually also the condition
under which the effects of thermal dissipation during inflation should be 
considered. 

Secondly, in the case of a thermal bath with a temperature higher than the Hawking
temperature, the thermal fluctuations of the scalar field plays an important and
even dominant role in producing the primordial perturbations of the universe.
Based on these results, the warm inflation scenario has been proposed. In
this model, the inflation epoch can smoothly evolve to a radiation-dominated epoch,
without the need of a reheating stage\cite{BF,LF}.  Dynamical analysis
of systems of inflaton with thermal dissipation\cite{OR} gives further
support to this model. It is found that the warm inflation solution is
very common.  A rate of dissipation as small as $10^{-7}\ H$, $H$ being the
Hubble parameter during inflation, can lead to a smooth exit from inflation
to radiation.

Warm inflation also provides explanation to the super-Hubble suppression.
The standard inflationary cosmology, which is characterized by an isentropic
de Sitter expansion, predicts that the particle horizon should be much larger
than the present-day Hubble radius $c/H_0$.  However, a spectral analysis of
the COBE-DMR 4-year sky maps seems to show a lack of power in the spectrum of
the primordial density perturbations on scales equal to or larger than the
Hubble radius $c/H_0$\cite{JF,BFH}. A possible explanation of this
super-Hubble suppression is given by hybrid models, where the primordial
density perturbations are not purely adiabatic, but mixed with an
isocurvature component. The warm inflation is one of the mechanisms which
can naturally produce both adiabatic and isocurvature initial
perturbations\cite{LF}.

In this paper, we study the power spectrum of mass density perturbations
caused by inflation with thermal dissipation. One purpose of developing the
model of warm inflation is to explain the amplitudes of the initial
perturbations.  Usually, the amplitude of initial perturbations from quantum
fluctuation of the inflaton depends on some unknown parameters of the
inflation potential. However, for the warm inflation model, the amplitude of
the initial perturbations is found to be mainly determined by the energy
scale
of inflation, $M$. If $M$ is taken to be about $10^{15}$ GeV, the possible
amplitudes of the initial perturbations are found to be in a range consistent
with the observations of the temperature fluctuations of the cosmic microwave
background (CMB) \cite{cobe}. That is, the thermally originated initial
perturbations apparently do not directly depend on the details of the
inflation potential, but only on some thermodynamical variables, such as the
energy scale $M$. This result is not unexpected, because like many
thermodynamical systems, the thermal properties including density
fluctuations should be determined by the thermodynamical conditions,
regardless of other details.

Obviously, it would be interesting to find more ``thermodynamical"  features
which contain only observable quantities and thermodynamical parameters, as
these predictions would be more useful for confronting models with
observations. Guided by these considerations, we will extend the
above-mentioned qualitative estimation of the order of the density
perturbations to a quantitative calculation of the power spectrum of the
density perturbations. We show that the power spectrum of the warm inflation
does not depend on unknown parameters of the inflaton potential and the
dissipation, but only on the energy scale $M$. The spectrum is found to be of
power law, and the index of the power law can be larger or less than 1.  More
interestingly, we find that for a given $M$, the amplitude and the index of
the power law are not independent from each other. In other words, the
amplitude of the power spectrum is completely determined by the power index
and the number $M$.  Comparing this result with the observed temperature
fluctuations of the CMB, we find that both amplitude and index of the power
spectrum can be fairly well fitted if $M \sim 10^{15}-10^{16}$ GeV. 

This paper is organized as follow: In Sec. II we discuss the evolution of the
radiation component for inflationary models with dissipation prescribed by a
field-dependent friction term. In particular, we scrutinize the physical
conditions on which the thermal fluctuations dominate the primordial density
perturbations.  Section III carries out the calculations of the power
spectrum of the density perturbations of the warm inflations. And finally, in
Sec. IV we give the conclusions and discuss further observational tests.

\section{Inflation with Thermal Dissipation}

\subsection{Basic equations}

Let us consider a flat universe consisting of a scalar inflaton field $\phi$
and a thermal bath. Its dynamics is described by the following
equations\cite{BF}. The equations of the expanding universe are
%eq2.1
\begin{equation}
2 \dot{H} +3 H^{2} = -\frac{8\pi}{m_{\rm Pl}^2}
\left[\frac{1}{2}\dot{\phi}^2 + \frac{1}{3}\rho_{r} - V(\phi)\right],
\end{equation}
%eq2.2
\begin{equation}
H^{2} = \frac{8\pi}{3}\frac{1}{m_{\rm Pl}^2}\left[\rho_{r} +
\frac{1}{2}\dot\phi^{2} + V(\phi)\right],
\end{equation}
where $H=\dot{R}/R$ is the Hubble parameter, and $m_{\rm Pl} = \sqrt {1/G}$
the Planck mass. $V(\phi)$ is the effective potential for field $\phi$, and
$\rho_{r}$ is the energy density of the thermal bath. Actually the scalar
field $\phi$ is not uniform due to fluctuations. Therefore, the field
$\phi$ in Eqs.(2.1) and (2.2) should be considered as an average over
the fluctuations.

The equation of motion for scalar field $\phi$ in a de-Sitter universe is
%eq2.3
\begin{equation}
\ddot{\phi} + 3H \dot{\phi} + \Gamma\dot{\phi} 
- e^{-2Ht}\nabla^2\phi+ V'(\phi)=0,
\end{equation}
where the friction term $\Gamma \dot{\phi}$ describes the interaction
between
the $\phi$ field and a heat bath. Obviously, for a uniformed field, or
averaged $\phi$, the term $\nabla^2\phi$ of Eq. (2.3) can be ignored.
Statistical mechanics of quantum open
systems has shown that the interaction of quantum fields with thermal or
quantum bath can be described by a general fluctuation-dissipation
relation\cite{weiss}. It is probably reasonable to describe the interaction
between the inflaton and the heat bath as a ``decay" of the inflaton 
\cite{BHP}. These results
support the idea of introducing a damping or friction term into the field
equation of motion. In particular, the friction term with the form in Eq.
(2.3), $\Gamma\dot{\phi}$, is a possible  approximation for the dissipation
of $\phi$ field in a heat bath environment in the near-equilibrium
circumstances. In principle, $\Gamma$ can be a function of $\phi$. In the
cases of polynomial interactions between $\phi$ field and bath environment,
one may take the polynomial of $\phi$ for $\Gamma$, {\it i.e.},
$\Gamma= \Gamma_m \phi^m$. The friction coefficient must be positive
definite, hence $\Gamma_m > 0$, and the dissipative index of friction $m$
should be zero or even integer if $V(\phi)$ is invariant under the
transformation $\phi \rightarrow -\phi$.

The equation of the radiation component (thermal bath) is given by the first
law  of thermodynamics as
%eq2.4
\begin{equation}
\dot \rho_{r} + 4H\rho_{r} = \Gamma\dot{\phi}^{2}.
\end{equation}
The temperature of the thermal bath can be calculated by $\rho_{r}=
(\pi^{2}/30) g_{\rm eff}T^{4}$, $g_{\rm eff}$ being the effective number of
degrees of freedom at temperature $T$. 

The warm inflation scenario is generally defined by a characteristic that
the thermal fluctuations of the scalar field dominate over the quantum
origin of the initial density perturbations. Because the thermal and quantum
fluctuations of the scalar field are proportional to $T$ and $H$
respectively, a necessary condition for warm inflation models is the
existence of a radiation component with temperature
%eq2.5
\begin{equation}
T > H
\end{equation}
during the inflationary expansion. Eq. (2.5) is also necessary for
maintaining the thermal equilibrium of the radiation component. In general,
the time scale for the relaxation of a radiation bath is shorter for higher
temperature. Accordingly, to have a relaxing time of the bath shorter than
the expansion of the universe, a temperature higher than $H$ is generally
needed.

As a consequence of Eq. (2.5), warm inflation scenario requires that the
solutions of Eqs. (2.1) - (2.4) should contain an inflation era, followed by
smooth transition to a radiation-dominated era. Dynamical system analysis also 
confirmed that for a massive scalar field $V(\phi) = \frac{1}{2}M^2\phi^2$, the
warm inflation solution of Eqs. (2.1) - (2.4) is very common. A smooth exit from
inflation to radiation era can be established even for a dissipation with $\Gamma$
as small as $10^{-7}H$\cite{OR}. A typical solution of warm inflation will be
given in next section. 

\subsection{Evolution of Radiation component during inflation}

Since warm inflation solution does not rely on a specific potential, we will
employ the popular $\phi^4$ potential commonly used for the ``new"  inflation
models. It is
%eq2.6
\begin{equation}
V(\phi)=\lambda(\phi^{2}-\sigma^{2})^{2}.
\end{equation}
To have slow-roll solutions, the potential should be flat enough, {\it i.e.},
$\lambda \leq (M/m_{\rm Pl})^4$, where $V(0)\equiv M^{4}=\lambda \sigma^{4}$.

For models based on the potential of Eq. (2.6), the existence of a thermal
component during inflation seems to be inevitable. In order to maintain the
$\phi$ field close to its minimum at the onset of the inflation phase
transition, a thermal force is generically necessary. In other words, there
is, at least, a weak coupling between $\phi$ field and other fields
contributing to the thermal bath. During the slow roll period of inflation,
the potential energy of the $\phi$ field is fairly constant, and their
kinetic energy is small, so that the interaction between the $\phi$ field
with the fields of the thermal bath remains about the same as at the
beginning. As such, there is no compelling reason to ignore these
interactions.

Strictly speaking, we should use a finite temperature effective potential
$V(\phi,T)$. However, the correction due to finite temperature is negligible. 
The leading temperature correction of the potential (2.6) is $\lambda T^2
\phi^2$. On the other hand, as mentioned above, we have $\lambda \leq
(M/m_{\rm Pl})^4$ for the flatness of the potential. Therefore, $\lambda T^2
\leq M^6/m_{\rm Pl}^4 \sim (M/m_{\rm Pl})^2H^2 \ll H^2$, {\it i.e.}, the
influence of the finite temperature effective potential can be ignored when
$\phi < m_{\rm Pl}$. 

Now, we try to find warm inflation solutions of Eqs. (2.1) - (2.4) for weak
friction $\Gamma < H$. In this case, Eqs. (2.1) - (2.3) are actually
the same as the ``standard" new inflation model when
%eq2.7
\begin{equation}
\rho_r \ll V(0).
\end{equation}
Namely, we have the slow-roll solution as
%eq2.8
\begin{equation}
\dot{\phi} \simeq - \frac{V'(\phi)}{3H + \Gamma(\phi)} \simeq
\frac{V'(\phi)}{3H},
\end{equation}
%eq2.9
\begin{equation}
\frac{1}{2} \dot{\phi}^2 \ll V(0),
\end{equation}
and
%eq2.10
\begin{equation}
H^2 \simeq H_i^2 \equiv \frac{8\pi}{3}\frac{V(0)}{m_{\rm Pl}^2}
\simeq \left(\frac{M}{m_{\rm Pl}}\right)^2 M^2,
\end{equation}
where the subscript $i$ denotes the starting time of the inflation epoch.

During the stage of $\phi \ll \sigma$, it is reasonable to neglect the
$\phi^{3}$ term in Eq. (2.3). We have then
%eq2.11
\begin{equation}
\ddot{\phi}+(3H + \Gamma)\dot{\phi}-4\lambda\sigma^2 \phi=0.
\end{equation}
Considering $\Gamma < H$, an approximate solution of $\phi$ can immediately
be found as
%eq2.12
\begin{equation}
\phi =\phi_i e^{\alpha Ht},
\end{equation}
where $\alpha \simeq \lambda^{1/2}(m_{\rm Pl}/M)^2/2\pi$ and $\phi_i$ is the
initial value of the scalar field. 

Substituting solution (2.12) into Eq. (2.4), we have the general solution
of (2.4) as
%eq2.13
\begin{equation}
\rho_r(t)  = A e^{(m+2)\alpha Ht} + B e^{-4Ht}
\end{equation}
where $A=\alpha^{2}H\Gamma_m\phi_i^{m+2}/[(m+2)\alpha+4]$, $B = \rho_r(0) -
A$, and $\rho_r(0)$ is the initial radiation density. Obviously, the term $B$
in Eq. (2.12) describes the blowing away of the initial radiation by the
inflationary exponential expansion, and the term $A$ is due to the generation
of radiation by the $\phi$ field decay. 

According to Eq. (2.13), the evolution of the radiation has two phases.
Phase
1 covers the period during which the $B$ term is dominant, and radiation
density drops drastically due to the inflationary expansion. The component of
radiation evolves into phase 2 when the $A$ term becomes dominant, where the
radiation density increases due to the friction of the $\phi$ field. Namely,
both heating and inflation are simultaneously underway in phase 2. Therefore,
this phase is actually the era of inflation plus reheating. 

The transition from phase 1 to phase 2 occurs at time $t_b$ determined by
$(d\rho/dt)_{t_b}=0$. We have
%eq2.14
\begin{equation}
Ht_b \simeq \frac{1}{(m+2)\alpha + 4}
\ln \left\{\frac{4[(m+2)\alpha+4]}{(m+2)\alpha^3H}\cdot
\frac {aM^4}{\Gamma_m\phi_i^{m+2}} \right\},
\end{equation}
where $a \equiv (\pi^2/30)g_{\rm eff}$. Then the radiation density at the
rebound time becomes
%eq2.15
\begin{equation}
\rho_r(t_b)= \frac{1}{4}
\left[(m+2)\alpha+4\right]A\exp [(m+2)\alpha Ht_b].
\end{equation}

 From Eqs. (2.12) and (2.13), the radiation density in phase 2 is given by
%eq2.16
\begin{equation}
\rho_r(t) = \frac{1}{4}\alpha^2 \Gamma H \phi^2(t)
\simeq \frac{1}{16\pi^2} \lambda\left(\frac{m_{\rm Pl}}{M} \right )^4
\Gamma H \phi^2(t).
\end{equation}
Since $H \simeq (M/m_{\rm Pl})M$, Eq. (2.16) can be rewritten as
%eq2.17
\begin{equation}
\rho_r(t) \sim
\lambda^{1/2} \left(\frac{m_{\rm Pl}}{M} \right )^2 \frac{\Gamma}{H}
\left (\frac{\phi(t)}{\sigma}\right)^2 V(0).
\end{equation}
On the other hand, from (2.12), we have
%eq2.18
\begin{equation}
\frac{1}{2}\dot{\phi}(t)^2 \simeq \lambda^{1/2}
\left(\frac{m_{\rm Pl}}{M} \right )^2
\left (\frac{\phi(t)}{\sigma}\right)^2 V(0).
\end{equation}
Therefore, in the case of weak dissipation $\Gamma <H$, we have
%eq2.19
\begin{equation}
\rho_r (t) <  \dot{\phi}(t)^2 /2.
\end{equation}
This is consistent with the condition of inflation
Eq. (2.7) when Eq. (2.9) holds. 

Eqs. (2.7) and (2.9) indicate that the inflation will come to an end at time
$t_f$ when the energy density of the radiation components, or the kinetic
energy of $\phi$ field, $\dot{\phi}^2/2$, become large enough, and
comparable to $V(0)$. From Eqs. (2.17) and (2.18), $t_f$ is given by
%eq2.20
\begin{equation}
\lambda^{1/2} \left(\frac{m_{\rm Pl}}{M} \right )^2
\left (\frac{\phi(t_f)}{\sigma}\right)^2 \simeq 1.
\end{equation}

In general, at the time when the phase 2 ends, or a radiation-dominated era
starts, the potential energy may not be fully exhausted yet. In this case, a
non-zero potential $V$ will remain in the radiation-dominated era, and the
process of $\phi$ decaying into light particles is still continuing. 

However, considering $\lambda^{1/2} (m_{\rm Pl}/{M})^2(\Gamma/H) <1$, the
right hand side of Eq. (2.17) will always be less than 1 when $\phi(t)$ is
less than $\sigma$. This means that, for weak dissipation, phase 2 cannot
terminate at $\phi(t) < \sigma$, or $V(\phi(t_f)) \neq 0$.  Therefore, under
weak dissipation, phase 2 will end at the time $t_f$ when the potential
energy $V(\phi)$ is completely exhausted, {\it i.e.},
%eq2.21
\begin{equation}
\phi(t_f) \sim \sigma.
\end{equation}
This means that no non-zero $V$ remains once the inflation exits to a
radiation-dominated era, and the heating of $\phi$ decay also ends at $t_f$. 

\subsection{Temperature of radiation}

 From Eq. (2.13), one can find the temperature $T$ of the radiation in phases 1
($t<t_b$) and 2 ($t>t_b$) as
%eq2.22
\begin{equation}
T(t)  = \left\{ \begin{array}{ll}
                T_{b}e^{-H(t-t_b)},   & \mbox{if \ $t < t_b$}, \\
       T_{b} e^{(m+2)\alpha H (t-t_b)/4}, & \mbox{if \ $t_f > t > t_b$},
                 \end{array}
        \right.
\end{equation}
where
%eq2.23
\begin{equation}
T_b = (4a)^{-1/4}[(m+2)\alpha+4]^{1/4}A^{1/4}
\exp \left[\left(\frac{m+2}{4}\right)\alpha Ht_b \right].
\end{equation}
The temperature $T_f$ at the end of phase 2 is
%eq2.24
\begin{equation}
T_f=T(t_f) = T_b e^{(m+2)\alpha H (t_f-t_b)/4},
\end{equation}
where $t_f$ is given by Eq. (2.21).

Since $T(t)$ is increasing with $t$ in phase 2, the condition (2.5) for warm
inflation can be satisfied if $T(t_f) > H$, or
%eq2.25
\begin{equation}
\rho_r(t_f) > a H_i^4.
\end{equation}
 Using Eqs. (2.17) and (2.21), condition (2.25) is realized if
%eq2.26
\begin{equation}
\frac{\Gamma}{H} > \left(\frac{\sigma}{m_{\rm Pl}}\right)^2
    \left(\frac{M}{m_{\rm Pl}} \right )^4.
\end{equation}
Namely, $\Gamma$ can be as small as $10^{-12} H$ for $M \sim 10^{16}$ GeV,
and
$\sigma \sim 10^{19}$ GeV. Therefore, the radiation solution (2.13), or warm
inflation, should be taken into account in a very wide range of dissipation
%eq2.27
\begin{equation}
10^{-12} H < \Gamma < H.
\end{equation}
This result is about the same as that given by dynamical system
analysis\cite{OR}: a tiny friction $\Gamma$ may lead the inflaton to a smooth
exit directly at the end of the inflation era. 

A typical solution of the evolution of radiation temperature $T(t)$ is
demonstrated in Fig. 1, for which parameters are taken to be $M = 10^{15}$
GeV, $\sigma = 2.24 \cdot 10^{19}$ GeV, $\Gamma_2 = 10^{-5}H_i$ and $g_{\rm
eff}$ = 100.  Actually, $g_{\rm eff}$-factor is a function of $T$ in general.
However, as can be seen below, the unknown function $g_{\rm eff}(T)$ has only
a slight effect on the problems under investigation. Figure 1 shows that the
rebound temperature $T_b$ can be less than $H$. In this case, the evolution
of $T(t)$ in phase 2 can be divided into two sectors:  $T < H$ for $t < t_e$,
and $T >H $ for $t > t_e$, where $t_e$ is defined by $T(t_e) =H$. We should
not consider the solution of radiation to be physical if $T < H$ since it is
impossible to maintain a thermalized heat bath with the radiation temperature
less than the Hawking temperature $H$ of an expanding universe. Nevertheless,
the solution (2.13) should be available if $t > t_e$. Therefore, one can
only consider the period of $t_e < t < t_f$ as the epoch of the warm
inflation.

Figure 1 also plots the Hubble parameter $H(t)$. The evolution of $H(t)$ is
about the same as in the standard new inflation model, {\it i.e.},
$H(t) \sim H_i$ in both phases 1 and 2. In Fig. 1, it is evident that the
inflation smoothly exits to a radiation era at $t_f$. The Hubble parameter
$H(t)$ also evolves from the inflation $H(t) \sim $ constant to a radiation
regime $H(t) \propto t^{-1}$.

The duration of the warm inflation is represented by ($t_f - t_e$) then. The
number of $e$-folding growth of the comoving scale factor $R$ during the warm
inflation is given by
%eq2.28
\begin{equation}
N \equiv \int_{t_e}^{t_f} Hdt \simeq \frac {4}{(m+2)\alpha} \ln \frac
{T_f}{H}.
\end{equation}
One can also formally calculate the number of $e$-folds of the growth in
phase  2 as
%eq2.29
\begin{equation}
N_2 \equiv \int_{t_b}^{t_f} H dt \simeq
\frac {4}{(m+2)\alpha} \ln \frac {T_f}{T_b},
\end{equation}
and the number of $e$-folds of the total growth as
%eq2.30
\begin{equation}
N_t \equiv \int_{0}^{t_f} Hdt \simeq \frac{4}{(m+2)\alpha}
\ln \left(\frac {T_f}{T_b}\right) + Ht_b.
\end{equation}

It can be found from Eqs. (2.28) - (2.30) that both $N_2$ and $N_t$ depend
on the initial value of the field $\phi_i$ via $T_b$, but $N$ does not. The
behavior of $T$ at the period $t > t_e$ is completely determined by the
competition between the diluting and producing radiation at $t>t_b$. Initial
information about the radiation has been washed out by the inflationary
expansion. Hence, the initial $\phi_i$ will not lead to uncertainty in our
analysis if we are only concerned the problems of warm evolution at the
period $t_e < t < t_f$.

\section{The primordial density perturbations}

\subsection{Density fluctuations of the $\phi$ field}

The fluctuations of $\phi$ field can be calculated by the similar way as 
stochastic inflations\cite{star}.  Recall that the coarse-grained scalar field
$\phi$ is actually determined from the decomposition between background and
high frequency modes, i.e. 
%eq3.1
\begin{equation}
\Phi({\bf x}, t)=\phi({\bf x}, t) + q({\bf x}, t),
\end{equation}
where $\Phi({\bf x},t)$ is the scalar field satisfying
%eq3.2
\begin{equation}
\ddot{\Phi} + 3H \dot{\Phi} - e^{-2Ht}\nabla^2\Phi+ V'(\Phi)=0.
\end{equation} 
$q({\bf x}, t)$ in Eq.(3.1) contains all high frequency modes and gives rise to
the thermal fluctuations. Since the mass of the field can be ignored for the high 
frequency modes, we have 
%eq3.3
\begin{equation}
q({\bf x}, t) = \int d^3k W(|{\bf k}|)
 \left [ a_{\bf k}\sigma_{\bf k}(t)e^{-i{\bf k}\cdot {\bf x}}
  + a^{\dagger}_{\bf k}\sigma^{*}_{\bf k}(t) e^{i{\bf k}\cdot {\bf x}} \right ]
\end{equation}    
where $k$ is comoving wave vector, and modes $\sigma_{\bf k}(t)$ is given by
%eq3.4
\begin{equation}
\sigma_{\bf k}(t) = \frac{1}{(2\pi)^{3/2}}\frac{1}{\sqrt{2k}}
\left [ H\tau - i\frac{H}{k}\right ] e^{-ik\tau},
\end{equation}
and $\tau=-H^{-1}\exp(-Ht)$ is the conformal time. Eq.(3.3) is appropriate in 
the sense that the self-coupling of the $\phi$ field is negligible. Considering
the high frequency
modes are mainly determined by the heat bath, this approximation is
reasonable. The window function  $W(|{\bf k}|)$ is properly chosen to filter 
out the modes at scales larger than the horizon size $H^{-1}$, {\it i.e.}, 
$W(k)=\theta(k-k_{h}(t))$, where $k_{h}(t)\simeq (1/ \pi)H \exp(Ht)$
\footnote{The coefficient $1/\pi$ actually depends on the details of the
cut-off function, which may not be step-function-like. For instance, 
considering causality, the cut-off function can be soft, and the 
longest wavelength of fluctuations can be a few times of the size 
of horizon\cite{BFH}} is the lower limit to the wavenumber of thermal fluctuations.

 From Eqs.(3.1) and (3.3), with the slow-roll condition, Eq.(3.2) renders
%eq3.5
\begin{equation}
3H\dot{\phi} - e^{-2Ht}\nabla^2\phi+ V'(\Phi)|_{\Phi=\phi}=
3H\eta({\bf x},t),
\end{equation}
and
%eq3.6
\begin{equation}
\eta({\bf x}, t)=\left (-\frac{\partial}{\partial t}
   +\frac{1}{3H}e^{-2Ht}\nabla^2\right) q({\bf x}, t).
\end{equation}
Eq.(3.5) can be rewritten as 
%eq3.7
\begin{equation}
\frac {d\phi({\bf x}, t)}{dt} =  - \frac{1}{3H}
\frac{\delta F[\phi({\bf x}, t)]}{\delta \phi} + 
\eta({\bf x}, t)
\end{equation}
where
%eq3.8
\begin{equation}
F[\bar{\phi}]=\int d^3{\bf x} \left[ \frac{1}{2} 
(e^{-Ht}\nabla\bar{\phi})^2 +
V(\bar{\phi}) \right]
\end{equation}
Eq. (3.7) is, in fact, the rate equation of the order parameter $\phi$ of 
a system with free energy $F[\phi]$. It describes the approach to equilibrium 
for the system during phase transition.

Using the expression of free energy (3.8), the slow-roll solution (2.8) can be 
rewritten as 
%eq3.9
\begin{equation}
\frac {d\phi}{dt} =  - \frac{1}{3H + \Gamma}
\frac{dF[\phi]}{d \phi}. 
\end{equation}
Hence, in the case of weak dissipation ($\Gamma < H$), Eq. (3.7) is 
essentially the same as the slow-roll solution (2.8) or Eq. (3.9) but with 
fluctuations $\eta$. The existence of the noise field ensures that the dynamical 
system properly approaches the global minimum of the inflaton potential 
$V(\phi)$. Strictly speaking, both the dissipation $\Gamma$ and fluctuations 
$\eta$ are consequences derived from $q({\bf x},t)$. They should be considered
together. 
However, it seems to be reasonable to calculate the fluctuations alone if 
the dissipation is weak.

Unlike (3.2), the Langevin equation (3.7) is of first order ($\dot{\phi}$) due
to the slow-roll condition. 
Generally, thermal fluctuations will cause both growing and decaying 
modes\cite{GM} \footnote{We thank the referee for pointing this problem out.}. 
Therefore, the slow-roll condition simplifies the problem from two types of 
fluctuation modes to one, {\it i.e.}, we can directly calculate the total
fluctuation as the superposition of various fluctuations. It 
has been shown \cite{berera}
that during the eras of dissipations, the growth of the structures in the
universe is substantially the same as surface roughening due to stochastic
noise. The evolution of the noise-induced surface roughening is described by
the so-called  KPZ-equation \cite{kardar}. Eqs.(3.5) or (3.7), which
includes terms of non-linear drift plus stochastic fluctuations, is a typical
KPZ-like equation.

 From Eq.(3.6), the two-point correlation function of $\eta({\bf x}, t)$
can be found as
%eq3.10
\begin{equation}
\langle \eta({\bf x}, t) \eta({\bf x'}, t') \rangle
=\frac{H^3}{4\pi^2}\left [ 1+ \frac{2}{\exp(H/\pi T)-1}\right]
\frac{\sin(k_{h}|{\bf x}-{\bf x'}|)}{k_{h}|{\bf x}-{\bf x'}|}
\delta(t-t'),
\end{equation}
where $1/[\exp(H/\pi T)-1]$ is the Bose factor at temperature $T$.
Therefore, when $T > H$, we have
%eq3.11
\begin{equation}
\langle \eta({\bf x}, t) \eta({\bf x}, t') \rangle
=\frac{H^2T}{2\pi}\delta(t-t').
\end{equation}
This result can also be directly obtained via the fluctuation-dissipation 
theorem\cite{hohe,hu}. In order to accord with the dissipation terms of 
Eq. (3.7), the fluctuation-dissipation theorem requires the ensemble average of
$\eta$ to be given by
%eq3.12
\begin{equation}
\langle \eta \rangle = 0
\end{equation}
%eq3.13
and
\begin{equation}
\langle \eta({\bf x},t) \eta({\bf x},t') \rangle = D\delta(t-t') \ .
\end{equation}
The variance $D$ is determined by
%eq3.14
\begin{equation}
D=2\frac{1}{U} \frac{T}{3H+\Gamma}~,
\end{equation}
where $U = (4\pi/3) H^{-3}$ is the volume with Hubble radius $H^{-1}$. In 
the case of weak dissipation, we then recover the same result as in Eq.(3.11),
%eq3.15
\begin{equation}
D=H^2T/2\pi.
\end{equation}
 When $T=H$, we obtain
%eq3.16
\begin{equation}
D=\frac{H^3}{2\pi},
\end{equation}
which agrees exactly the result derived from quantum fluctuations of
$\phi$-field\cite{star}. Therefore, the quantum fluctuations of
inflationary $\phi$ field are equivalent to the thermal noises stimulated by
a thermal bath with the Hawking temperature $H$. Eqs. (3.15) and (3.16) show
that the condition (2.5) is necessary and sufficient for a warm inflation.

For long-wavelength modes, the $V'(\phi)$ term is not negligible.
It may lead to a suppression of correlations on scales larger than 
$|V''(\phi)|^{-1/2}$. However, before the inflaton actually rolls down to the
global minimum, we have $|V''(\phi\ll \sigma)|^{-1/2} \geq H^{-1}$. 
The so-called abnormal dissipation of density perturbations \cite{kirk} may 
produce more longer correlation time than $H$. Therefore in phase 2, {\it i.e.},
the warm inflation phase $H< T < M$, the long-wavelength suppression will not 
substantially change the scenario presented above.

The fluctuations $\delta \phi$ of the $\phi$ field can be found
from linearizing Eq. (3.7). If we only consider the fluctuations
$\delta \phi$ crossing outside the horizon, {\it i.e.}, with wavelength 
$\sim H^{-1}$, the equation of $\delta \phi$ is
%eq3.17
\begin{equation}
\frac{d \delta \phi}{dt}= - 
\frac{H^2 + V^{''}(\phi)}{3H+\Gamma}\delta \phi + \eta.
\end{equation}
For the slow-roll evolution, we have $|V^{''}(\phi)| \ll 9H^2$ \cite{KT}.
One can ignore the $V^{''}(\phi)$ term on the right hand side of Eq. (3.17).
Accordingly, the correlation function of the fluctuations is
%eq3.18
\begin{equation}
\langle \delta \phi(t) \delta \phi(t') \rangle
\simeq D \frac{3H+\Gamma}{2H^2} e^{-(t-t')H^2/(3H +\Gamma)}, 
\ \ \ t > t',
\end{equation}
hence
%eq3.19
\begin{equation}
\langle (\delta \phi)^2 \rangle  
\sim \frac{3}{4\pi}HT
\end{equation}
Thus, in the period $t_e< t < t_f$ the density perturbations on large scales
are produced by the thermal fluctuations that leave the horizon with a
Gaussian-distributed amplitude having a root-mean-square dispersion given by
Eq. (3.19).

Principally, the problem of horizon crossing of thermal fluctuations
given by Eq. (3.7)  is different from the case of quantum
fluctuations, because the equations of $H$ and $\dot H$, (2.1) and (2.2)
contain terms in $\rho_r$. However, these terms are insignificant for
weak dissipation [Eq. (2.19)] in phase 2. Thus Eqs.(2.1) and (2.2) depend
only nominally on the evolution of $\rho_r$. Accordingly, for weak
dissipation, the behavior of thermal fluctuations at horizon crossing can
be treated by the same way as the evolutions of quantum fluctuations in
stochastic inflation. In that theory, quantum fluctuations of inflaton are
assumed to become classical upon horizon crossing and act as stochastic
forces. Obviously, this assumption is not necessary for thermal fluctuations. Moreover,
we will show that in phase 2 the thermal stochastic force $HT$ is 
contingent upon the comoving scale of perturbations by a power law [Eqs. (2.21)
and (3.21)], and therefore the power spectrum of the thermal fluctuations obeys the
power law. This make it more easier to estimate the constraint quantity in the
super-horizon regime.

Accordingly, the density perturbations at the horizon re-entry epoch are
characterized by\cite{KT}
%eq3.20
\begin{equation}
\left(\frac{\delta\rho}{\rho} \right)_h =
\frac{-\delta\phi V^{\prime}(\phi)}{\dot\phi^{2} + (4/3)\rho_{r}} \ .
\end{equation}
All quantities in the right-hand side of Eq. (3.20) are calculated at the
time when the relevant perturbations cut across the horizon at the
inflationary epoch.

Using the solutions of $\phi$ and $\rho_r$ of warm inflation (2.12) and
(2.13), Eq. (3.20) gives
%eq3.21
\begin{equation}
\left(\frac{\delta\rho}{\rho} \right )_h  \simeq
\left(\frac{5\cdot3^{3m/2+4}}{2^{m+3}\cdot\pi^{m/2+3}}\right)^{\frac{1}{m+2}}
\cdot \left(\frac{\gamma_m}{g_{\rm eff}\alpha^m}\right)^{\frac{1}{m+2}}
\left(\frac{T}{H}\right)^{\frac{1}{2}\left(\frac{m-6}{m+2}\right)},
\end{equation}
where the dimensionless parameter $\gamma_m \equiv \Gamma_m H^{m-1}$, and $T$
is the temperature at the time when the considered perturbations $\delta
\rho_r$ crossing out of the horizon $H^{-1} \sim H_i^{-1}$. Eq. (3.21) shows
that the density perturbations are insensitive to the $g_{\rm eff}$-factor. 

\subsection{Power law index}

Since inflation is immediately followed by the radiation dominated epoch, the
comoving scale of a perturbation with crossing over (the Hubble radius)  at
time $t$ is given by
%eq3.22
\begin{equation}
\frac {k}{H_0} = 2\pi \frac{H}{H_0} \frac{T_0}{T_f}e^{H(t-t_f)},
\end{equation}
where $T_0$ and $H_{0}$ are the present CMB temperature and Hubble constant
respectively. Eq. (3.22) shows that the smaller $t$ is, the smaller $k$ will
be. This is the so-called ``first out - last in" of the evolution of density
perturbations produced by the inflation. 

Using Eqs. (2.22) and (3.22), the perturbations (3.21) can be rewritten as
%eq3.23
\begin{equation}
\left\langle \left ( \frac{\delta\rho}{\rho} \right )^2 \right \rangle_h
       \propto k^{(m-6)\alpha/4},  \ \ \ \ {\rm if} \ \ \ k > k_e,
\end{equation}
where $k_e$ is the wavenumber of perturbations crossing out of horizon
at $t_e$. It is
%eq3.24
\begin{equation}
k_e = 2\pi H\frac{T_0}{T_f} e^{H(t_e-t_f)}
\simeq 2\pi H\frac{T_0}{T_f} e^{-N}.
\end{equation}
Therefore, the primordial density perturbations produced during warm inflation
are of power law with an index $(m-6)\alpha/4$.  We may also express the power
spectrum of the density perturbations at a given time $t$. It is
%eq3.25
\begin{equation}
\left\langle \left ( \frac{\delta\rho}{\rho} \right )^2 \right \rangle_t
       \propto k^{3+n},  \ \ \ \ {\rm if} \ \ \ k > k_e,
\end{equation}
where the spectral index $n$ is
%eq3.26
\begin{equation}
n= 1 + \left(\frac{m-6}{4}\right)\alpha.
\end{equation}

Clearly, for $m = 6$, the warm inflation model generates a flat power spectrum
$n=1$, yet the power spectrums will be tilted for $m \neq 6$.  The dissipation
models $\Gamma = \Gamma_{m}\phi^{m}$ may not be realistic for higher $m$, but
we will treat $m$ like a free parameter in order to show that the results we
concerned actually are not very sensitive to these parameters. 

The warm inflation scenario requires that all perturbations on comoving scales
equal to or less than the present Hubble radius originate in the period of
warm inflation. Hence, the longest wavelength of the perturbations (3.24),
{\it i.e.}, $2\pi/k_{e}$, should be larger than the present Hubble radius
$H_0^{-1}$. We have then
%eq3.27
\begin{equation}
N > \ln \left(\frac{HT_0}{H_0T_f}\right) =
\ln \left(\frac{T_0}{H_0}\right) - \ln \left(\frac{T_f}{H}\right) \sim 55,
\end{equation}
where we have used $(T_0/H_0)\gg (T_f/H)$, as $T_f \leq M$. Using Eq. (2.28),
the condition (3.27) gives an upper bound to $\alpha$ for a given $m$ as
%eq3.28
\begin{equation}
\alpha_{\rm max} = \left(\frac{4}{m+2}\right) \frac
{\ln(T_f/H)}{\ln(T_0/H_0)}.
\end{equation}
Thus, the possible area of the index $n$ can be found from Eq. (3.27) as
%eq3.29
\begin{equation}
n = \left \{ \begin{array}{ll}
       1 - (6-m)\alpha_{\rm max}/4 \ {\rm to}\ 1, & \mbox{if \ $m<6$}, \\
       1\  {\rm to}\ 1 + (m - 6)\alpha_{\rm max}/4, & \mbox{if \ $m>6$}.
             \end{array}
\right.
\end{equation}
Therefore, the power spectrum is positive-titled ({\it i.e.}, $n>1$) if $m>6$, and
negative-titled ($n<1$) if $m < 6$. Figure 2 plots the allowed area of $n$ as
a function of the inflation mass scale $M$. Apparently, for $M \geq 10^{16}$
GeV, the tilt $|n-1|$ should not be larger than about 0.15 regardless of the
values of $m$ from 2 to 12. 

\subsection{Amplitudes of perturbations}

To calculate the amplitude of the perturbations we rewrite spectrum (3.25) 
into
%eq3.30
\begin{equation}
\left\langle \left ( \frac{\delta\rho}{\rho} \right )^2 \right \rangle_h
      = A \left( \frac{k}{k_0}\right )^{n-1},  \ \ \ \ {\rm if} \ \ \ k > k_e,
\end{equation}
where $k_0=2\pi H_0$. $A$ is the spectrum amplitude normalized on scale
$k=k_0$, corresponding to the scale on which the perturbations re-enter the
Hubble radius $1/H_0$ at present time. From Eqs. (3.21), and (3.23), we have
%eq3.31
\begin{equation}
A =
\left(\frac{5\cdot3^{3m/2+4}}{2^{m+3}\cdot\pi^{m/2+3}}\right)^{\frac{2}{m+2}}
\left(\frac{\gamma_m}{g_{\rm eff}\alpha^m}\right)^{\frac{2}
    {m+2}}\left(\frac{H_0T_f}{HT_0}\right)^{n-1}
    \left(\frac{T}{H}\right)^{\frac{m-6}{m+2}} e^{(n-1)H(t_f - t)}.
\end{equation}
Applying Eq. (2.21), the radiation temperature at the moment of
horizon-crossing, $t$, can be expressed as $T(t) = T_f\exp[(m+2)\alpha
H(t-t_f)/4]$.  With the help of Eq. (2.28), we obtain
%eq3.32
\begin{equation}
\left(\frac{T}{H}\right)^{\frac{m-6}{m+2}} \left(\frac{T_f}{H}\right)^{n-1}
e^{(n-1)H(t_f-t)} = \exp\left\{(n-1)\left[1+\left(\frac{m+2}
{4}\alpha \right)\right]N\right\}.
\end{equation}
On the other hand, using Eqs. (2.20), (2.23) and (2.28), one has
%eq3.33
\begin{equation}
\gamma_m = \left(\frac{3}{4}\right)^{1-\frac{m}{2}}\cdot \frac{g_{\rm eff}}
{30}\left(\frac{M}{m_{\rm Pl}}\right)^{2m}\alpha^{-3-\frac{m}{2}}.
\end{equation}
Substituting Eqs. (3.32) and (3.33) into Eq. (3.31), we have finally
%eq3.34
\begin{equation}
A = \left(\frac{3^{4-m}}{64\pi^{3+\frac{m}{2}}}\right)^{\frac{2}{m+2}}
    \cdot \left(\frac{M}{m_{\rm Pl}}\right)^{\frac{4m}{m+2}}
    \left(\frac{H_0}{T_0}\right)^{n-1} \alpha^{-3} \exp \left\{(n-1)
    \left[1+\left(\frac{m+2}{4}\right)\alpha\right]N\right\}.
\end{equation}

Eq. (3.34) shows that the amplitude $A$ does not contain the unknown $g_{\rm
eff}$-factor. Moreover, $\alpha$ can be expressed by $n$ and $m$ through Eq.
(3.26), and $N$ can be expressed by $\alpha$ and $M$ via Eq. (2.28).
Therefore, the amplitude of the initial density perturbations, $A$, is only a
function of $M$, $n$, and $m$. 

Figures 3 and 4 plot the relations between the amplitude $A$ and index $n$ for
various parameters $M$ and $m$. In the case of $m=6$, $n=1$, the relation of
$A$ and $\alpha$ is plotted in Fig. 5. It can be seen from Figs. 3, 4 and 5
that for either $m \geq 6$ or $m < 6$, the amplitude $A$ is significantly
dependent on $M$, but not so sensitive to $m$. Namely, the testable $A$-$n$
relationship is mainly determined by a thermodynamical variable, the energy
scale $M$. This is a ``thermodynamical" feature. The relationship between $A$
and $N$ plotted in Figs. 6 and 7 also show this kind of ``thermodynamical" 
feature: the $A$-$N$ relation depends mainly on $M$. 

For comparison, the observed results of $A$ and $n$ derived from the 4-year
COBE-DMR data (quadrupole moment $Q_{rms-PS} \sim 15.3_{-2.8}^{+3.7}\mu K$ and
$n \sim 1.2 \pm 0.3$\cite{cobe}) are plotted in Figs. 3, 4 and 5. The
observationally allowed $A$-$n$ range is generally in a good agreement with
the predicted $A$-$n$ curve if $M \sim 10^{15} -10^{16}$ GeV, regardless the
parameter $m$. Figures 3 and 4 also indicate that if the tilt of spectrum
$|n-1|$ is larger than 0.1, the parameter area of $M\leq 10^{14}$ GeV will be
ruled out. Therefore, the warm inflation seems to fairly well reconcile the
initial perturbations with the energy scale of the inflation. 

\section{Conclusions and Discussion}

Assuming that the inflaton $\phi$-field undergoes a dissipative process with
$\Gamma\dot{\phi}^2$, we have studied the power spectrum of the mass density
perturbations. In this analysis, we have employed the popular $\phi^4$
potential. However, only one parameter, the mass scale of the inflation $M$,
is found to be important in predicting the observable features of power
spectrum, {\it i.e.}, the amplitude $A$ and index $n$. Actually, the warm
inflation scenario is based on two thermodynamical requirements: (a) the
existence of a thermalized heat bath during inflation, and (b) that the
initial fluctuations are given by the fluctuation-dissipation theorem. 
Therefore, we believe that the ``thermodynamical" features -- $A$ and $n$
depend only on $M$ -- would be generic for the warm inflation.  This feature
is useful for model testing. Hence, the warm inflation can be employed as
an effective working model when more precise data about the observable
quantities $A$, $n$ etc. become available.  The current observed data of $A$
and $n$ from CMB are consistent with the warm inflation scenario if the mass
scale $M$ of the inflation is in the range of $10^{15} - 10^{16}$ GeV.

\acknowledgments
We would like to thank an anonymous referee for a detailed report that
improved the presentation of the paper.  Wolung Lee would like to thank Hung 
Jung Lu for helpful discussions.

%\newpage

\begin{figure}

\caption{A typical solutions of the evolutions of $\phi$ field and radiations
in warm inflation in which $\Gamma = \Gamma_2 \phi^2$, $V(\phi) =
\lambda(\phi^2 - \sigma^2)^2$ and $\lambda \sigma^4 = M^4$, $M$ being the
inflaton energy scale. The parameters are taken to be $M = 10^{15}$GeV,
$\sigma = 2.24 \cdot 10^{19}$ GeV, $\Gamma_2 = 10^{-5}H_i$ and $g_{\rm eff}$ =
100.  The dot-dashed and solid lines are for $H(t)$ and the radiation
temperature $T$ respectively.  $t_b$ is the time at which the temperature
rebound ($T_b$) and $t_e$ the time of $T = H$.  The inflation ends at $t_f$
when the temperature is $T_f$.  $T$ and $H$ are in units of $H_i \equiv [8\pi
V(0)/3m_{\rm Pl}^2]^{1/2}$, and $t$ is in units of $H_i^{-1}$.}
\label{1}

\end{figure}

\begin{figure}

\caption{The allowed area of power law index $n$ as a function of the mass
scale $M$ for various $m$. For given $M$ and $m$, the possible $n$ should lie
between the line $n=1$ and the corresponding curve of $m$.  For $m = 6$, the
only solution is $n = 1$.} 
\label{2}

\end{figure}

\begin{figure}

\caption{The amplitudes of the power spectrum as a function of $n$ in the area
of $n<1$. The mass scales $M$ are labeled at the curves. The dotted, dashed,
and solid lines are for $m = 0, \ 2$ and 4 respectively.  All curves end at
the points when the corresponding warm inflation durations $N$ are less than
55.  The region within the dot-dashed box is the allowed area of ($n, A$) 
given by the 4-year COBE-DMR data.} 
\label{3}

\end{figure}

\begin{figure}

\caption{The same as Fig. 3, but for $n>1$.  The dotted, dashed, and solid
lines are for $m = 8, \ 10$ and 12 respectively.} 
\label{4}

\end{figure}

\begin{figure}

\caption{The relation $A$ and $\alpha$ for $m = 6$. In this case, $n = 1$. The
dot-dashed line represents the COBE-DMR data at $n = 1$ and $Q_{rms-PS} = 15.3
\mu K$, {\it i.e.}, $A \simeq 3.5\times 10^{-6}$.} 
\label{5}

\end{figure}

\begin{figure}

\caption{The amplitudes of the power spectrum as a function of the thermal
duration $N$ at three inflaton mass scales $M$ and in the range of $n < 1$. 
The dotted, dotted and dashed lines are for $m = 0,\ 2$ and 4 respectively.}
\label{6}

\end{figure}

\begin{figure}

\caption{The same as Fig. 6, but for $n \geq 1$. The dotted, dotted and dashed
lines are for $m = 8,\ 10$ and 12 respectively.} 
\label{7}

\end{figure}

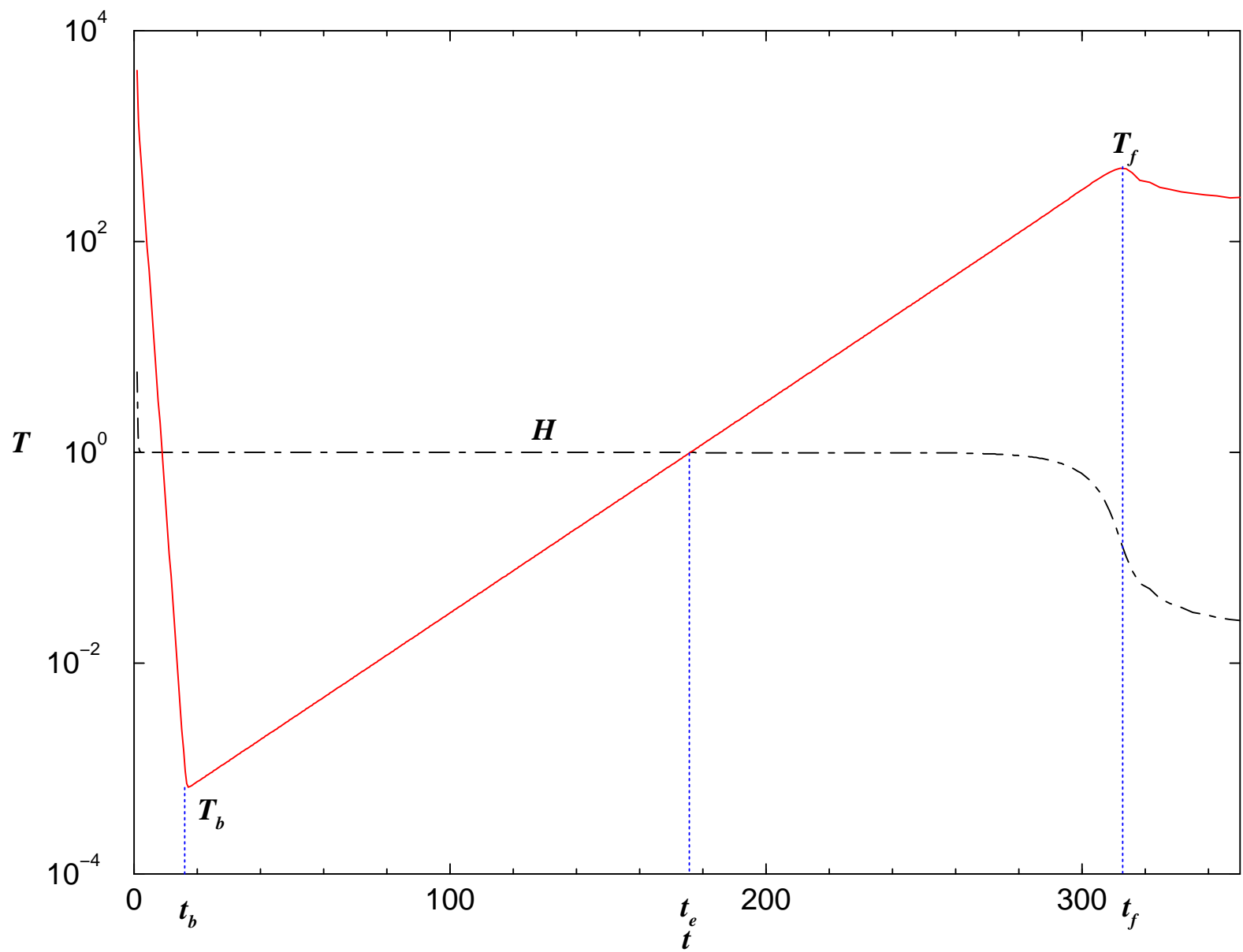
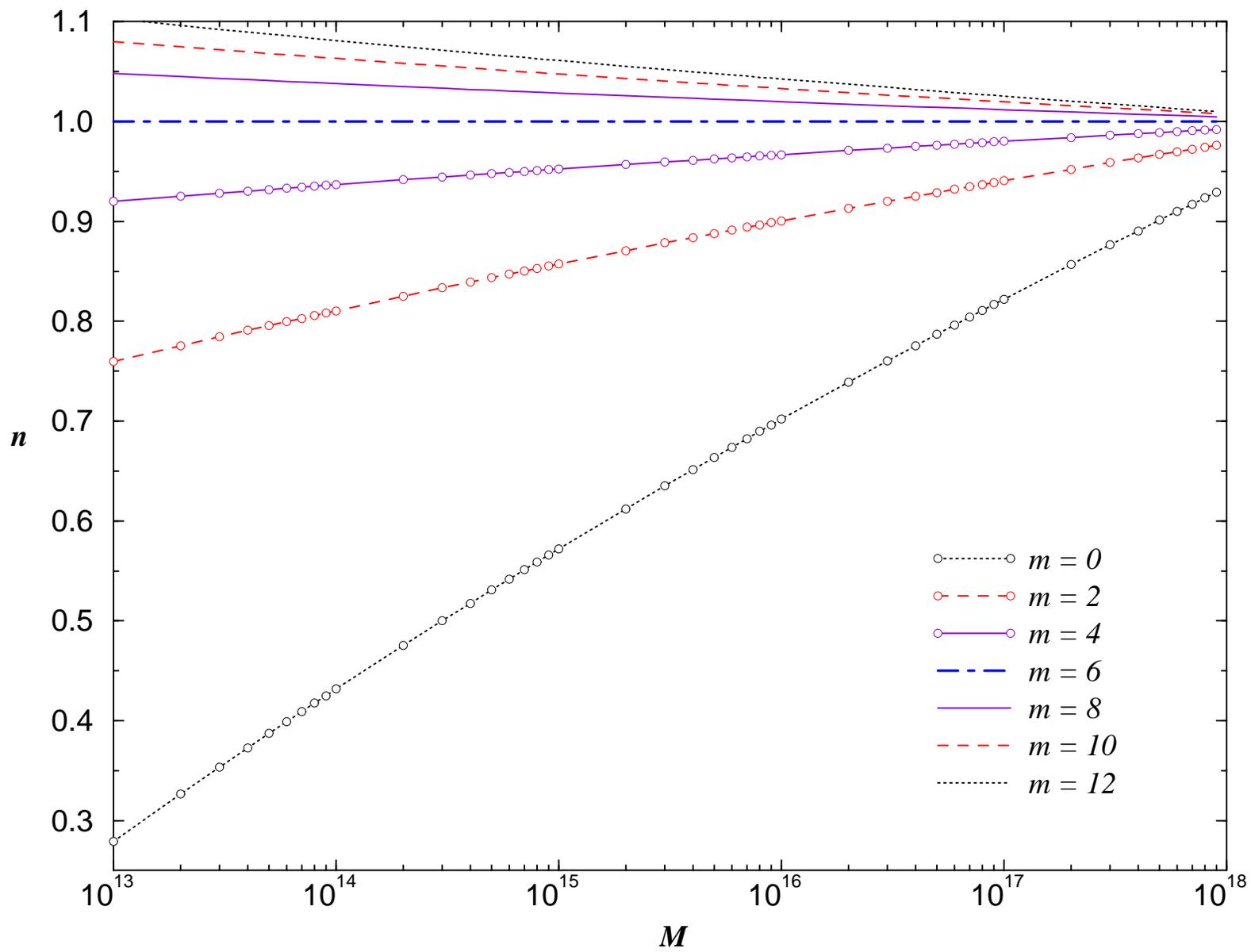
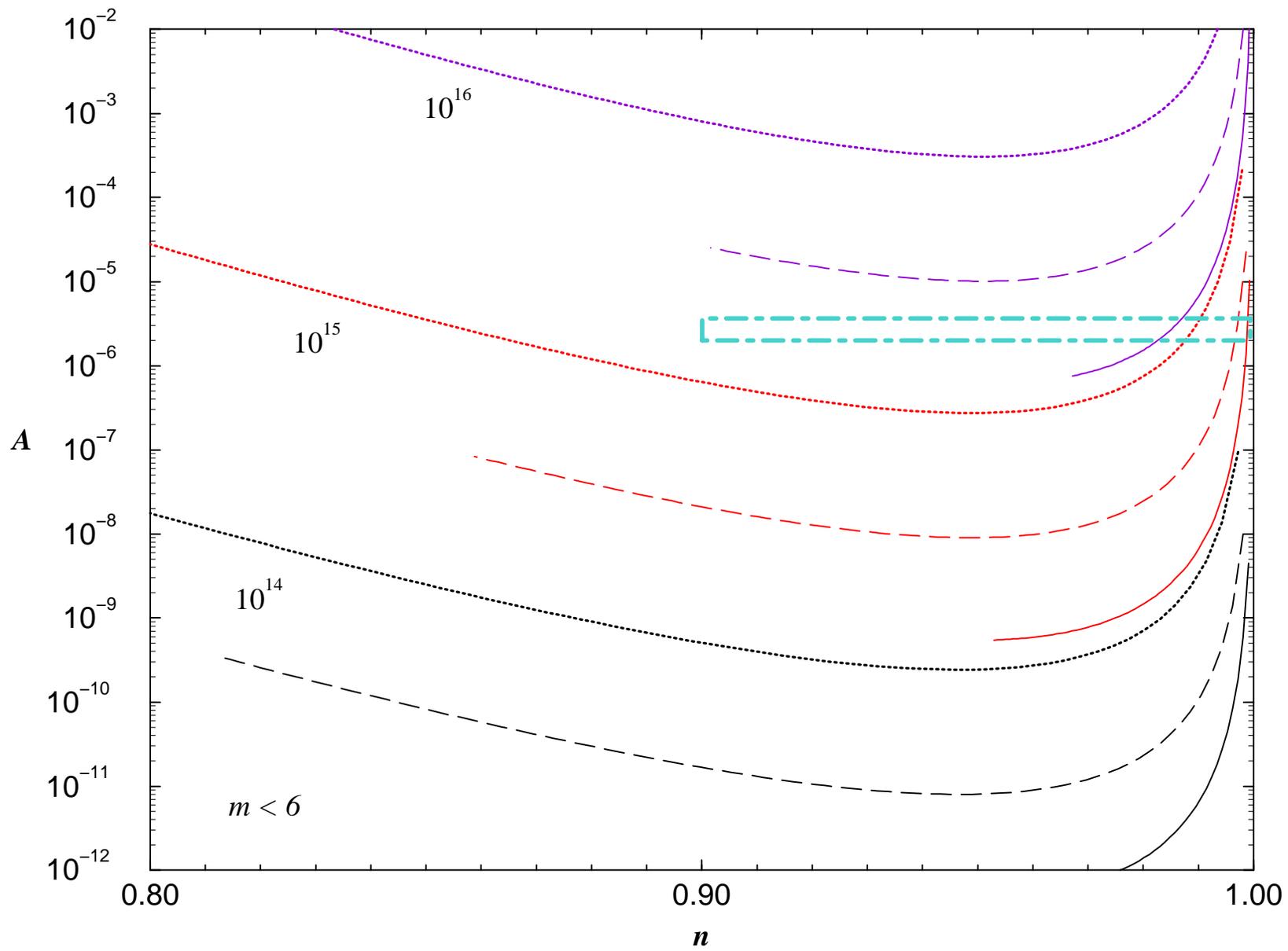
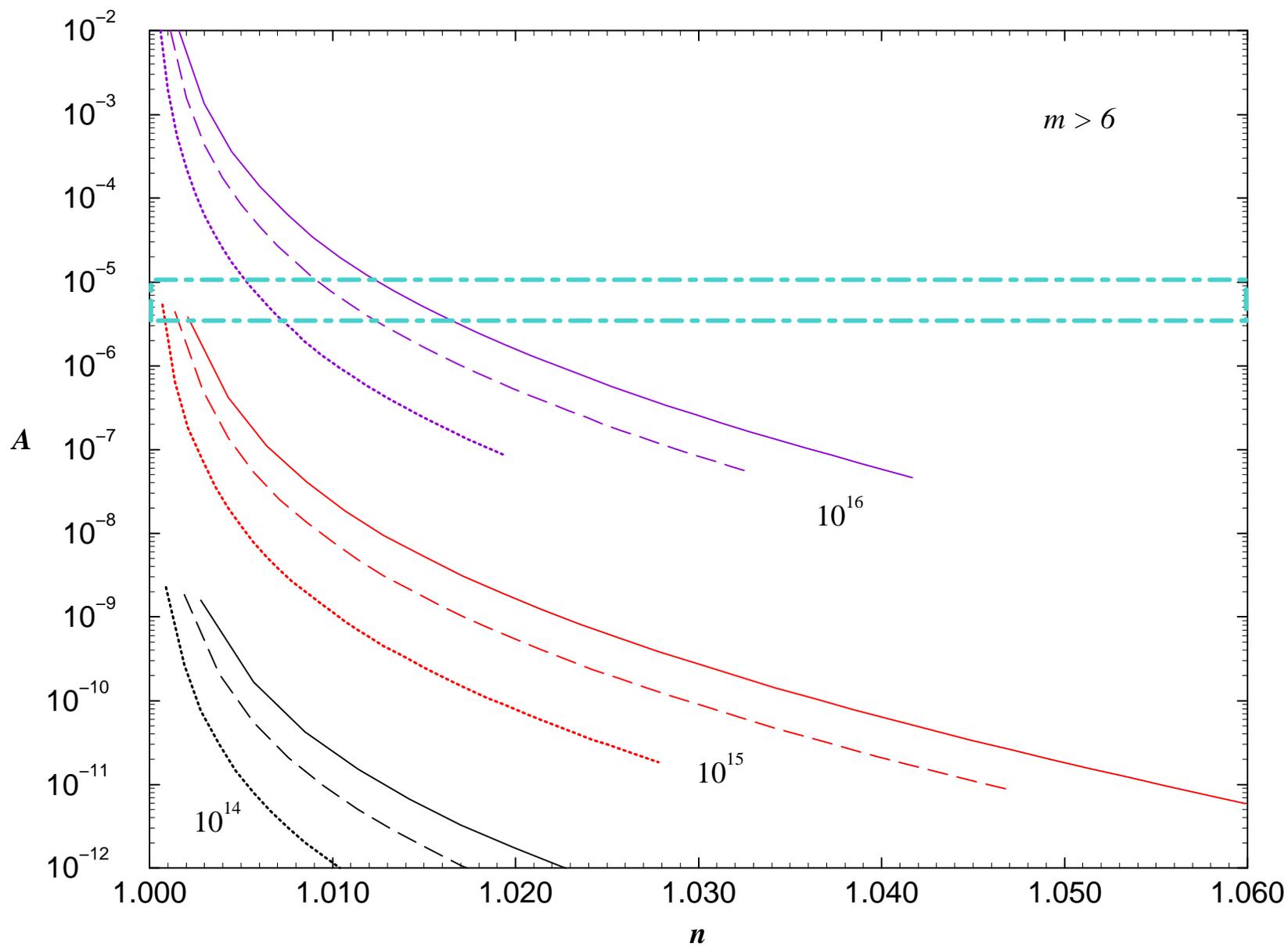
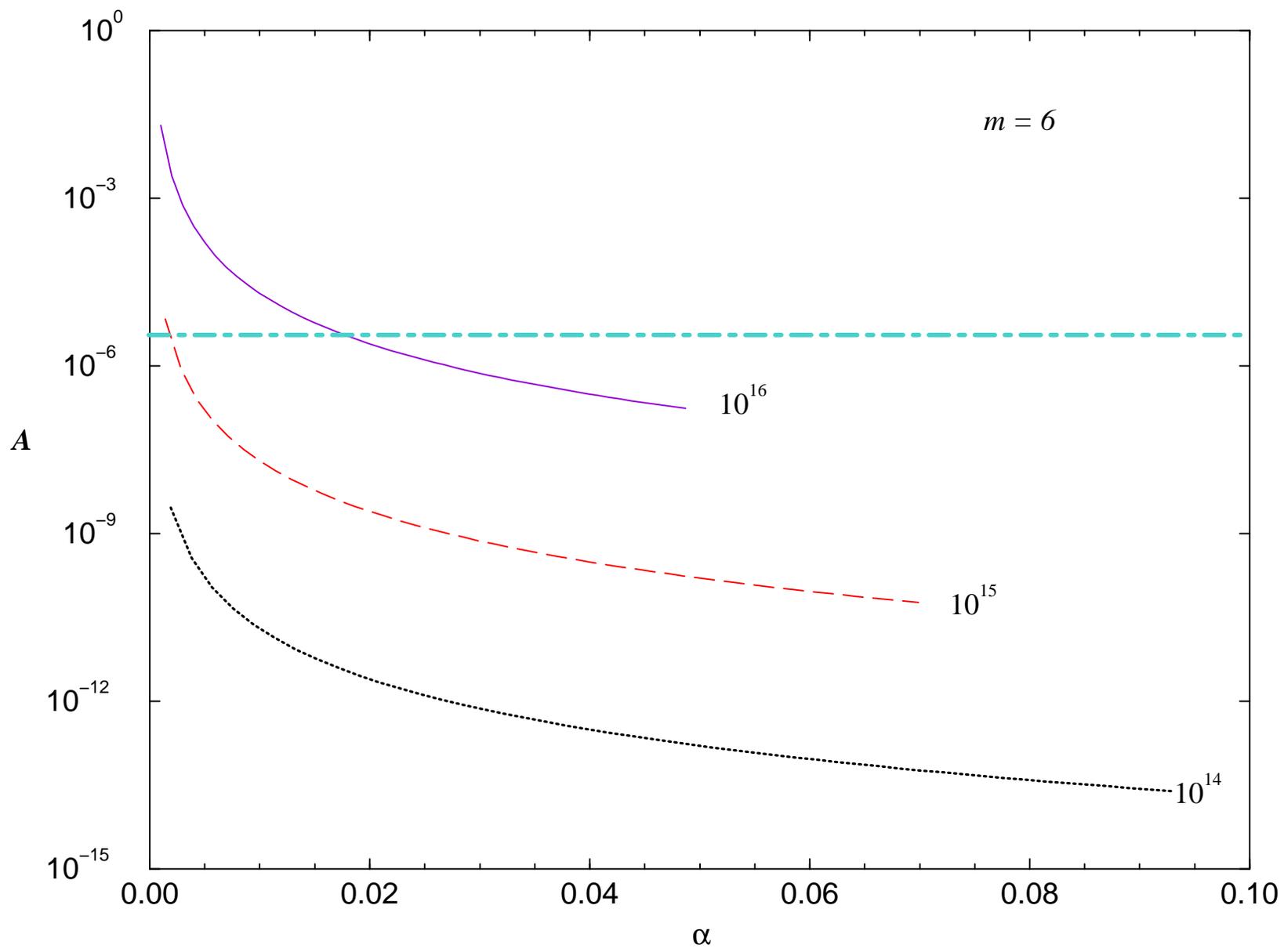
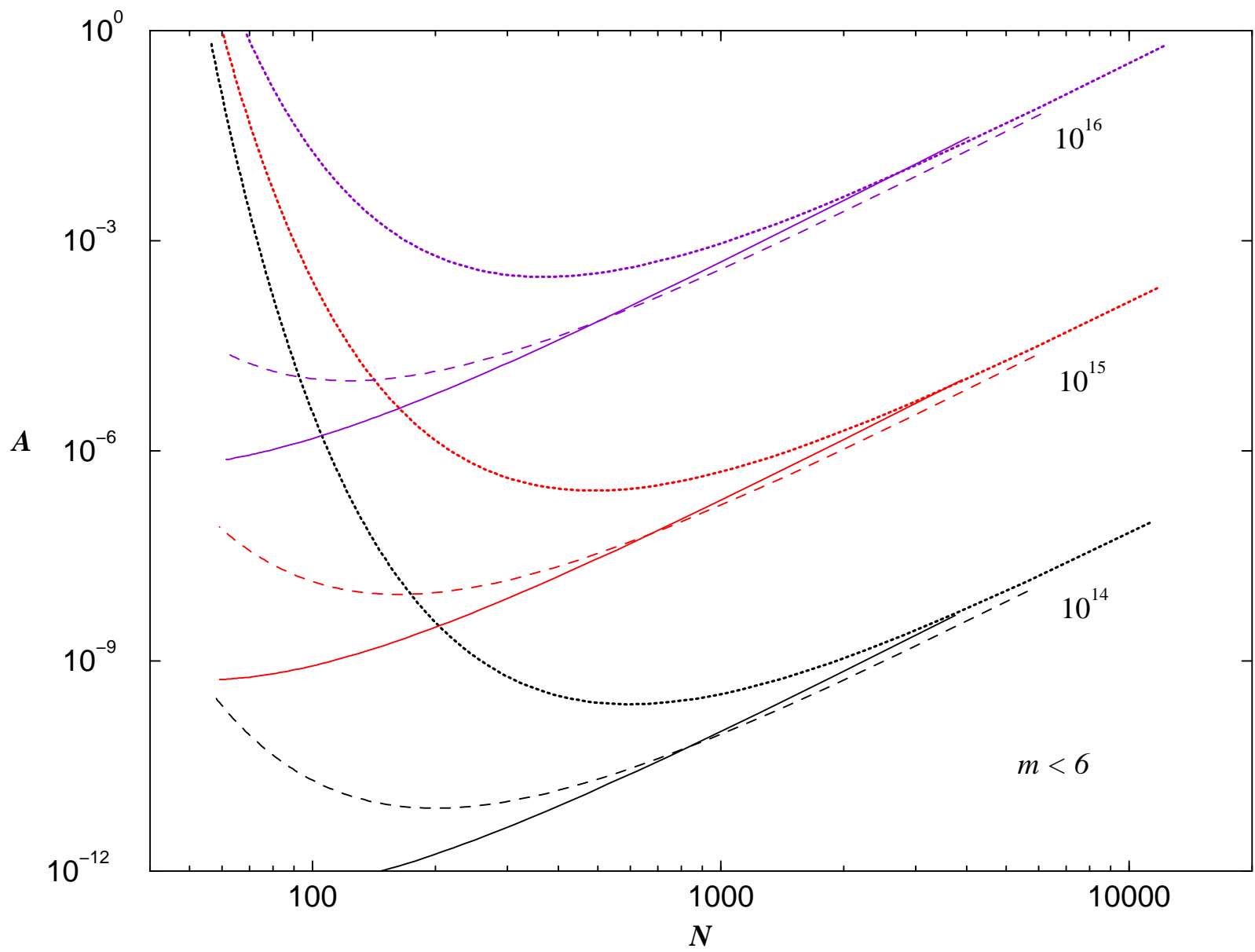
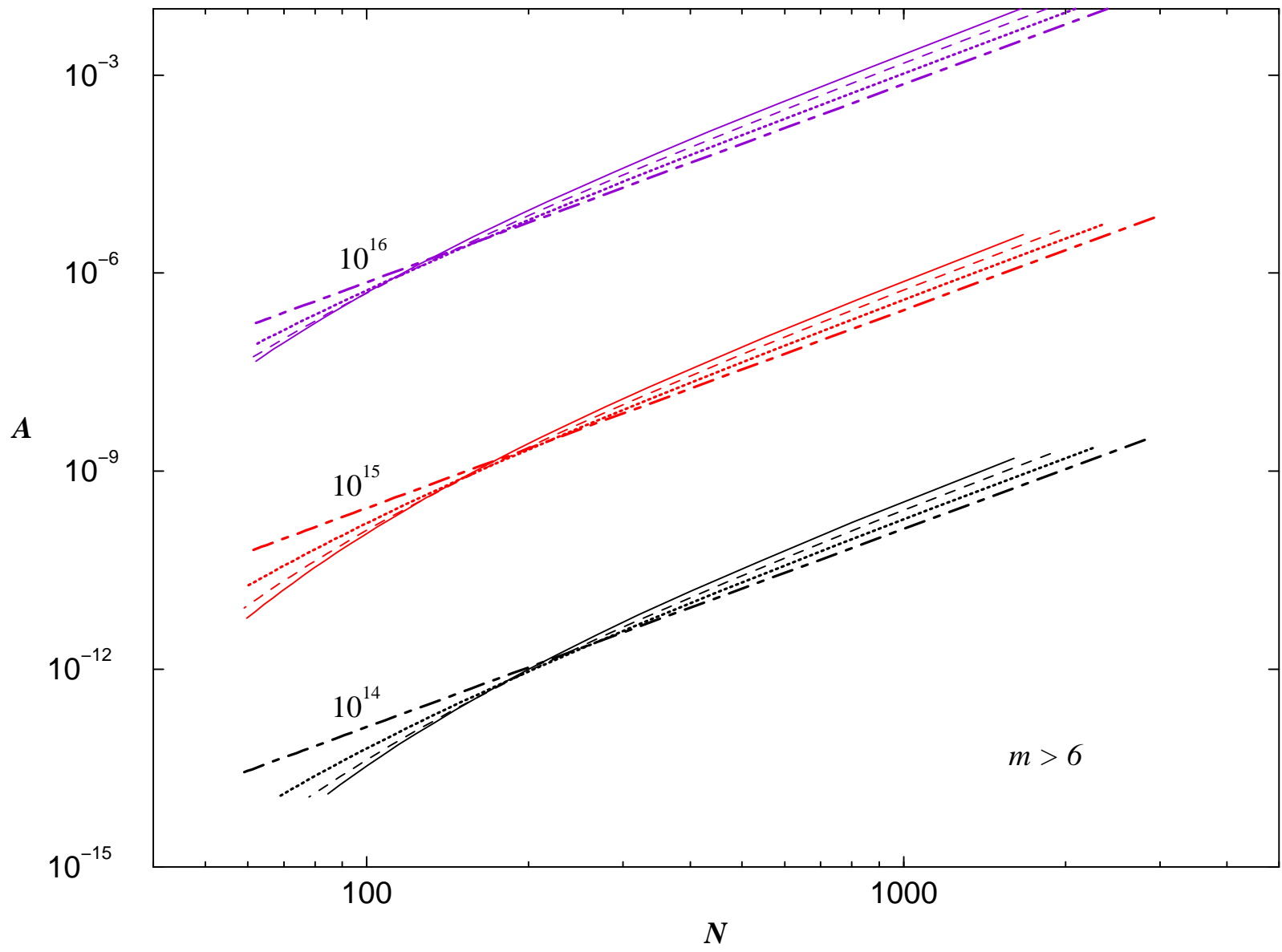
\end{document}